\documentclass[twocolumn,showpacs,preprintnumbers,amsmath,amssymb]{revtex4}
\usepackage{graphicx}
\usepackage{bm} 
\usepackage[]{amsfonts, graphicx, amsmath}
\DeclareGraphicsExtensions{.eps,.ps,.eps.gz,.ps.gz,.eps.Z}

\newcommand{\ra}{\rangle}
\newcommand{\la}{\langle}
\newcommand{\tr}{{\rm Tr}}

\begin{document}

\title{Hitting time for quantum walks on the hypercube}
\author{\sc Hari Krovi}\email{Email:  krovi@usc.edu}
\author{Todd Brun}\email{Email:  tbrun@usc.edu}
\affiliation{Communication Sciences Institute, University of Southern California, \\
Los Angeles, California 90089, USA}
\date{\today}

\begin{abstract}
Hitting times for discrete quantum walks on graphs give an average
time before the walk reaches an ending condition.  To be analogous to the
hitting time for a classical walk, the quantum hitting time must involve
repeated measurements as well as unitary evolution. We derive an expression
for hitting time using superoperators, and numerically evaluate it for the
discrete walk on the hypercube. The values found are compared to other analogues
of hitting time suggested in earlier work. The dependence of hitting
times on the type of unitary ``coin'' is examined, and we give an example of
an initial state and coin which gives an {\it infinite} hitting time for
a quantum walk. Such infinite hitting times require destructive interference,
and are not observed classically. Finally, we look at distortions of the hypercube,
and observe that a loss of symmetry in the hypercube increases the hitting time.
Symmetry seems to play an important role in both dramatic speed-ups and
slow-downs of quantum walks.
\end{abstract}

\maketitle

\section{Introduction}

Algorithms based on random walks have been immensely successful in
computer science. The most efficient solution for 3-SAT is based
on the hitting time of a classical random walk \cite{Mot}. They
are also useful in algorithms for k-SAT and probability
amplification. It therefore seems reasonable that their quantum
analogues might be useful in developing quantum algorithms. This
has been one of the main motivations for defining and analyzing
quantum walks. Most quantum algorithms to date
are based on the Quantum Fourier Transform (QFT), like Shor's
algorithms for factoring and discrete log \cite{Shor}. But
the use of QFT seems to be restricted. For example, it does not
seem to be useful in some non-Abelian hidden subgroup problems
like the Graph Isomorphism problem. Therefore, there is a need for
new algorithmic tools for the analysis of such problems. This is
another reason to study the properties of quantum walks on graphs.

Quantum walks come in two versions: continuous time and discrete
time. Continuous time versions have been studied in \cite{farhi},
\cite{childs}, \cite{childs2}. In \cite{childs2}, it has been
demonstrated that for the "glued trees" graph, the quantum walk is
exponentially faster than a classical random walk. Discrete time
versions have been studied extensively for the walk on the line
\cite{nayakV}, \cite{bachWatrous}, \cite{todd1}, \cite{todd2} and
the hypercube \cite{kempe_comp}, \cite{shenvi},
\cite{mooreRussell}. Quantum walks for any general (irregular)
graphs have been defined in \cite{ambainisReview} and \cite{viv}.
In \cite{shenvi}, it was shown that there is a quadratic speed up
with discrete quantum walks in searching an unstructured database
for a marked node. \cite{ambainisApp} describes how these concepts
can be used for the problem of element distinctness. Reviews on
the recent developments in this field can be found in
\cite{ambainisReview} and \cite{kempeReview}.

Classical random walks are characterized by a number of time scales:
mixing time, cover time, correlation time, and hitting time.
Similar quantities can be defined for quantum walks, analogous to
their classical counterparts, though some types of classical behavior
are not guaranteed to hold for quantum walks.
In \cite{aharanov} a lower bound on the mixing
time for quantum walks was found in terms of the conductance
of the underlying graph. Some notions of hitting time have been
defined in \cite{kempe_comp}, where an upper bound on the hitting
time has been found for the quantum walk on the hypercube.

In a classical discrete random walk, a {\it hitting time} is the average time
for a walk beginning on a particular starting vertex to arrive for the first time
at a particular ending vertex or group of vertices.  It is not sufficient
for the ending vertex merely to have positive probability:  the particle
must actually arrive there.  Therefore, hitting times are generally
longer than the shortest path between the starting and ending
vertices.  An analogous definition can be made for a continuous
random walk, but in this paper we will only consider discrete walks.

This definition becomes complicated for a quantum walk, where the
particle can be in a superposition of many vertex locations at once.
To capture the idea of first arrival time, our definition of
hitting time is the expected time for the particle to be at the final vertex
for the first time as determined by a measurement of the final vertex.  At every time step
we perform a measurement to see if the particle is at the final
vertex or not. This notion of hitting time seems to be the most
natural extension of the definition of hitting time to the quantum
case.  Using this definition, we will obtain an expression for hitting time on
any graph, and produce numerical results showing that it is orders of
magnitude lower than the classical hitting time for the quantum walk
on the hypercube.

This paper is organized as follows. In section II, a quantum walk on a
regular graph is defined and this definition applied to a walk
on the hypercube. In section III, we review the classical definition of
hitting time.  Then we define the quantum hitting time and
derive an expression for it on any regular graph (or on an
irregular graph when the walk is appropriately defined). We also briefly
review other definitions of hitting time that have been used for quantum
walks. In section IV, we discuss the results of simulations for a walk
on the hypercube using the Grover coin, and compare them to the
classical hitting time and the other definitions for a quantum
hitting time, showing that in this case the hitting time is far shorter than these other
definitions. Then in section V we show that the hitting time for a quantum
walk can be much longer than the classical hitting time for certain combinations
of initial state and coin, and can even be infinite.  We demonstrate this for a walk
on the hypercube with the discrete Fourier transform (DFT) coin.  In section
VI we look at a distorted version of the hypercube, and show that the loss
of symmetry increases the hitting time.  We conclude in section VII with a
discussion of the role of symmetry in the speed-up of quantum walks.

\section{Quantum walk on the hypercube}

\subsection{Quantum walks}

A discrete time quantum walk is defined as the application of a
unitary operator $\hat{U}$ on a Hilbert space representing the graph;
each vertex of the graph has a set of basis states associated with it,
and the unitary operator can only cause a transition between two
basis states if the vertices associated with them are connected by an edge.
If the graph is regular, the Hilbert space can be a tensor product of two Hilbert
spaces:  in a walk on a regular graph with $N$ vertices and degree
$d$, $\hat{U}$ is applied to an $Nd$-dimensional Hilbert space
$\mathcal{H}^p\bigotimes\mathcal{H}^c$, where $\mathcal{H}^p$ is the Hilbert
space of the position (i.e., the vertex), and $\mathcal{H}^c$ is the Hilbert
space of a $d$-sided ``coin,'' which is ``flipped" at each step to determine
which edge to walk on.

For a given graph, if we label each edge
incident on a vertex by a number from $1$ through $d$, such that
an edge connecting two vertices gets the same label at both ends,
then each vertex is a basis state in the position space and each
direction is a basis state in the coin space. Therefore, the basis
states for the position and coin spaces can be labelled
$|0\ra,|2\ra,\dots,|N-1\ra$ and $|0\ra,|2\ra,\dots,|d-1\ra$. The
unitary $\hat{U}$ is of the form
\begin{equation}
\hat{U} = \hat{S}\hat{F} ,
\label{evolution_op}
\end{equation}
where we call $\hat{S}$ 
the shift matrix and $\hat{F}$ the coin flip matrix. $\hat{S}$ is a permutation
matrix which shifts the particle from its present vertex along the edge
indicated by the coin state, in a way
analogous to a classical random walk.  The coin flip matrix $\hat{F}$
acts solely on the coin space, so it is of the form $\hat{I}\otimes\hat{C}$.
$\hat{C}$ can, in principle, be any unitary matrix, though in general we
concentrate on examples with some kind of useful structure.
The most common coins analyzed are the Grover coin $\hat{G}$,
the Discrete Fourier Transform (DFT) coin $\hat{D}$, and the Hadamard
coin $\hat{H}$.  While the Grover and the DFT coins exist for all
dimensions, the Hadamard coin exists only for dimensions $2^n$ for
some $n$. These coins are given by the following matrices:
\begin{equation}
\hat{G}=2|\Psi\ra\la\Psi|-I,
\end{equation}
where $|\Psi\ra=\frac{1}{\surd{d}}\sum_i|i\ra$, i.e.,
\begin{equation}
\hat{G}=\begin{pmatrix}
  \frac{2}{d}-1 & \frac{2}{d} & \ldots & \frac{2}{d} \\
  \frac{2}{d} & \frac{2}{d}-1 & \ldots & \frac{2}{d} \\
  \vdots & \ddots & \ddots & \vdots \\
  \frac{2}{d} & \frac{2}{d} & \ldots & \frac{2}{d}-1
\end{pmatrix} ,
\label{Grover_matrix}
\end{equation}
\begin{equation}
\hat{D}=\frac{1}{\surd{d}}\begin{pmatrix}
  1 & 1 & 1 & \ldots & 1 \\
  1 & \omega & \omega^2 & \ldots & \omega^{d-1} \\
  \vdots & \ddots &  & \ddots & \vdots \\
  1 & \omega^{d-1} & \omega^{2(d-1)} & \ldots &
  \omega^{(d-1)(d-1)}
\end{pmatrix} ,
\end{equation}
where $\omega=\exp(2\pi i/d)$, and
\begin{equation}
\hat{H}_{2^d} = \hat{H}_{2^{d-1}}\otimes \hat{H}_2 ,
\end{equation}
where
\begin{equation}
\hat{H}_2=\frac{1}{\surd{2}}\begin{pmatrix}
  1 & 1 \\
  1 & -1
\end{pmatrix}
\end{equation}
The shift operator is applied after the coin operator. The shift
operator moves the particle from a vertex along the edge given by
the direction number of the coin state i.e.,
\begin{equation}
\hat{S}=\sum_v\sum_i|v(i),i\ra\la v,i|,
\end{equation}
where $v(i)$ is the vertex connected to $v$ via the edge numbered
$i$. It is interesting to note that the term "quantum random walk"
is, in a sense, a misnomer, because the randomness in a quantum walk is
introduced by quantum measurements where one of the measurement
outcomes takes place at random. Thus, there is no extra randomness
in the walk, i.e., outside of what is introduced by quantum
mechanics itself.

\subsection{The Hypercube}

The hypercube of dimension $n$ is a set of $2^n$ vertices each
with a degree $n$. These vertices can be numbered by a $n$-bit
string from $(00\dots0)$ through $(11\dots1)$. Each edge leads in a
particular direction, and can be labeled by an integer $j$, $0\le j < n$.
Adjacent vertices are the ones whose bit assignments differ by a
single bit.  For example, for the hypercube in
3 dimensions, the vertices $(011)$ and $(111)$ are adjacent to
each other, connected by an edge in the 2 direction.  We can treat
the bit strings of the vertices as $n$-dimensional Boolean vectors,
whose elements are all 0 or 1; the directions of the edges then
correspond to vectors $\vec{e}_j$ with element $j$ equal to 1 and
all other elements zero.  The vertex adjacent to a given vertex 
$|\vec{x}\ra$ in the $j$ direction can be labeled
$|\vec{x}\oplus\vec{e_i}\ra$ $i\epsilon\{0,1,\dots,n-1\}$.

Since the vertices of the
hypercube can be labeled $|0\ra$ through $|2^n-1\ra$ and the
directions labeled $|0\ra$ through $|n-1\ra$, the Hilbert space
is $\mathcal{H}^{2^n}\bigotimes\mathcal{H}^n$.  The shift
operator is of the form
\begin{equation}
\hat{S} = \sum_{j=0}^{n-1}\sum_{\vec{x}}|\vec{x}\oplus\vec{e}_j,j\ra\la
\vec{x},j|,
\end{equation}
where $|v,j\ra=|v\ra\otimes |j\ra$, $|v\ra$ is the position state
and $|j\ra$ the coin state.

\section{Hitting time}

\subsection{Classical hitting time}

Given a regular undirected graph and a particle which starts at
some vertex, the classical random walk is defined as follows.  At
each vertex, the particle moves along any edge incident on the
vertex with some predefined probability.  This procedure is then
repeated at the new vertex.  The walk continues until the particle
arrives at (``hits'') a certain vertex (called the ``final vertex'') for
the first time.  The {\it hitting time} is defined as the average
time until the particle hits the final vertex:
\begin{equation}
\tau(v) = \sum_{t=0}^\infty t p_v(t) ,
\end{equation}
where $\tau(v)$ is the hitting time given that the walk starts at
vertex $v$ and $p_v(t)$ is the probability that the particle hits
the final vertex for the first time at time step $t$ (first crossing
probability) given that it was at $v$ at $t=0$.

Let us now specialize to the case of the hypercube, where the
the final vertex is assumed to be $11\cdots1$.  We would like to
find the hitting time starting from $00\cdots0$.
For the classical walk on the hypercube, one can
arrive at a recursive relation involving the hitting time. First,
from the symmetry of the hypercube one can conclude that the
hitting time depends only on the {\it hamming weight} of the starting vertex
rather than the vertex itself.  The hamming weight is the number of
1's in the string of bits.  At hamming weight $x$, there are
$C^n_x=n!/x!(n-x)!$ vertices.  The probability to walk
to a vertex with weight $x+1$ is $(n-x)/n$, and the probability
to walk to a vertex with weight $x-1$ is $x/n$.
So, if $\tau(x)$ denotes the hitting time
starting at any vertex with hamming weight $x$, then
\begin{equation}
\tau(x)=\frac{n-x}{n}\tau(x+1)+\frac{x}{n}\tau(x-1)+1,
\end{equation}
with the boundary condition $\tau(n)=0$.  This simplifies to
\begin{equation}
\Delta(x)=\frac{n-x-1}{x+1}\Delta(x+1)-\frac{n}{x+1},
\end{equation}
where $\Delta(x)=\tau(x)-\tau(x+1)$. Using this recursive formula,
we obtain
\begin{equation}
\tau(0)=\sum_{x=0}^{n-1}\Delta(x)=\sum_{x=0}^{n-1}\frac{\sum_{j=0}^{x-1}C_{x-j}^n+1}{C_x^{n-1}},
\end{equation}
This sum can readily be evaluated for reasonable sizes of $n$;  we use
this expression to compare the classical hitting time to the
quantum hitting time. We define the hitting time of a quantum walk next.

\subsection{Quantum hitting time}

We define the hitting time of a quantum walk in close analogy to
that of a classical random walk.  For a classical walk, hitting time
is the average time taken for the particle to hit the
final vertex for the first time.  To carry this over to
the quantum case, we must give a proper meaning to the phrase ``for the
first time.'' The only reasonable way to do this is to measure the final vertex at every step to
see if the particle has arrived or not. Without
such a measurement at every step, one cannot appropriately define
the first crossing probability.
To define the hitting time this way, we first define
what is called a measured quantum walk on a graph.

Suppose we have a walk on a regular graph of degree d with N
vertices, where the particle begins at a vertex $x_0$ and walks
till it reaches the vertex $x_f$. Assume that the coin starts in
the initial state $|c_0\rangle$. The initial state of the particle
is then,$|\Psi\rangle=|x_0\rangle\otimes|c_0\rangle$;
$\rho_0=|\Psi\rangle\langle\Psi|$ where $\rho_0$ is the density
operator corresponding to the initial state $|\Psi\ra$. The
measured walk is now the alternating application of a unitary evolution
operator $\hat{U}$ (the product of the shift and the coin flip
operators) and a projective measurement $M$ with 2
outcomes $\hat{P}_f$ and $\hat{Q}_f = \hat{I} - \hat{P}_f$,
where $\hat{P}_f = |x_f\ra\la x_f|\otimes\hat{I}$
is the projector onto the final vertex for any coin state.
Thus, if the particle is not found in the final vertex after $t$ time steps, the state is:
\begin{equation}
\rho_t = \frac{ (\hat{Q}_f\hat{U})^t\rho((\hat{Q}_f\hat{U})^\dag)^t }{ \tr\{ (\hat{Q}_f\hat{U})^t\rho((\hat{Q}_f\hat{U})^\dag)^t \} }.
\end{equation}
If the particle {\it is} found in the final vertex, the walk is assumed to end.
The state $|x_f\rangle$ becomes an absorbing boundary for
this (measured) walk. Now, the first crossing probability at time
step $t$ can be defined (analogous to the classical case) to be the
following expression:
\begin{equation}\label{probEqn}
p(t)=\tr\{\hat{P}_f\hat{U}[\hat{Q}_f\hat{U}]^{t-1}
\rho_0[\hat{U^{\dag}}\hat{Q}_f]^{t-1}\hat{U^{\dag}}\hat{P}_f\} .
\end{equation}
With this definition for the probability, the hitting time becomes
\begin{equation}\label{ht}
\tau = \sum_{t=1}^\infty t p(t).
\end{equation}
To sum this series requires a slightly different expression for the first arrival
probability $p(t)$.  We rewrite (\ref{probEqn}) in terms
of the following {\it superoperators} (i.e., linear transformations on operators):
\begin{eqnarray}
\mathcal{N}\rho = \hat{Q}_f\hat{U}\rho\hat{U^{\dag}}\hat{Q}_f \nonumber\\
\mathcal{Y}\rho=\hat{P}_f\hat{U}\rho\hat{U^{\dag}}\hat{P}_f.
\end{eqnarray}
In terms of $\mathcal{N}$ and $\mathcal{Y}$,
$p(t)=Tr\{\mathcal{Y}\mathcal{N}^{t-1}\rho_0\}$.
We now evaluate the hitting time by introducing a new superoperator,
\begin{equation}
\mathcal{O}(l) = l\sum_{t=1}^\infty (l\mathcal{N})^{t-1} ,
\label{superop_sum}
\end{equation}
which is a function of a parameter $l$.  The hitting time now becomes
\begin{equation}
\tau=\frac{d}{dl}Tr\{\mathcal{Y}\mathcal{O}(l)\rho_0\}\biggr|_{l=1}.
\label{derivative_form}
\end{equation}
If the superoperator
$\mathcal{I}-l\mathcal{N}$ is invertible, then we can replace the
sum (\ref{superop_sum}) with the closed form
\begin{equation}
\mathcal{O}(l) = l(\mathcal{I}-l\mathcal{N})^{-1}.
\end{equation}
If $\mathcal{I}-l\mathcal{N}$ is {\it not} invertible, then this
sum instead is a pseudoinverse---that is, the inverse restricted to the support
of the superoperator.  So long as the graph in question is finite,
this is well-defined.  The derivative in (\ref{derivative_form}) is then
\begin{equation}
\frac{d\mathcal{O}}{dt}(1)
= (\mathcal{I}-\mathcal{N})^{-1}+\mathcal{N}(\mathcal{I}-\mathcal{N})^{-2}
= (\mathcal{I}-\mathcal{N})^{-2}.
\end{equation}
This gives us the following expression for the hitting time:
\begin{equation}
\tau = \tr\{\mathcal{Y}(\mathcal{I}-\mathcal{N})^{-2}\rho_0\}.
\label{closed_form_tau}
\end{equation}
The meaning for the terms in the above expression (e.g.
$\mathcal{I}-\mathcal{N})^{-2}$) can be given by {\it vectorizing} all
the operators.  Any matrix can be vectorized by turning its rows into columns and
stacking them up one by one, so that a $D\times D$ matrix becomes
a column vector of size $D^2$.  For example:
\begin{equation}
\begin{pmatrix}
a_{11} & a_{12} & a_{13} \\
a_{21} & a_{22} & a_{23} \\
a_{31} & a_{32} & a_{33}
\end{pmatrix}
\rightarrow
\begin{pmatrix}
a_{11} \\
a_{12} \\
a_{13} \\
a_{21} \\
a_{22} \\
a_{23} \\
a_{31} \\
a_{32} \\
a_{33}
\end{pmatrix} . \nonumber
\end{equation}
Consequently the superoperators become matrices
of size $D^2\times D^2$.

Denoting the vectorized quantities as $\rho^v$, etc., we obtain the
hitting time as
\begin{equation}
\tau=\tr^v\{\mathcal{Y}(\mathcal{I}-\mathcal{N})^{-2}\rho^v\} ,
\end{equation}
where
$(\mathcal{Y}\rho)^v=(\mathcal{\hat{P}}_f\hat{U}\rho\hat{U^{\dag}}\mathcal{\hat{P}}_f)^v$
and
$(\mathcal{N}\rho)^v=(\mathcal{\hat{Q}}_f\hat{U}\rho\hat{U^{\dag}}\mathcal{\hat{Q}}_f)^v$
and $\tr^v\{.\}$ is the equivalent of the trace operation for vectorized
quantities. It is just the inner product of the resultant vector
with $I^v$, the vectorized identity matrix:  $\tr^v\{ O^v \} = I^v \cdot O^v$.

Using Roth's lemma \cite{roth}, we obtain
\begin{eqnarray}
(\mathcal{N}\rho)^v &=& \left[ (\hat{Q}_f\hat{U})\otimes(\hat{Q}_f\hat{U})^\ast \right] \rho^v ,
\nonumber \\
(\mathcal{Y}\rho)^v &=& \left[ (\hat{P}_f\hat{U})\otimes(\hat{P}_f\hat{U})^\ast \right] \rho^v .
\end{eqnarray}
Therefore, the expression for hitting time becomes
\begin{equation}\label{hittime}
\tau = I^v \cdot \left( (\hat{P}_f\hat{U})\otimes(\hat{P}_f\hat{U})^\ast(\mathcal{I}-
(\hat{Q}_f\hat{U})\otimes(\hat{Q}_f\hat{U})^\ast)^{-2}\rho^v \right) .
\end{equation}

\subsection{Other definitions of hitting time}

Two definitions of the quantum hitting time were given in
\cite{kempe_comp}:  {\it one-shot hitting time} and {\it concurrent hitting
time}. The one-shot hitting time is defined for an unmeasured walk.
It is the time at which the probability of being in the final
state is greater than some given value. More precisely, given some
probability $p$, the one-shot hitting time is defined as the lowest time
$T$ such that
\begin{equation}
|\la x_f|\hat{U}^T|x_0\ra|^2\geq p
\end{equation}
where $x_f$ and $x_0$ are the final and initial states, and
$\hat{U}$ is the evolution operator (as defined above).
Essentially this same definition of hitting time was
used in the analysis of the continuous-time walk on the hypercube in \cite{Alagic05}.
This definition is useful if it is known that at some time the probability to be
in the final state will be higher than some reasonable value; but for a general graph,
this is not guaranteed.

The concurrent hitting time, by contrast, is defined for a measured quantum
walk. Given a probability $p$, it is the the time $T$ such that
the measured walk has a probability greater than $p$ of stopping
at a time less than $T$. It has been proved that the concurrent
hitting time is $T=\frac{\pi}{2}$ for $p=\Omega(\frac{1}{nlog^2n})$ for a
hypercube of dimension $n$.
Since we consider only the measured quantum walk in this paper, we
compare our numerical results to the numerical simulation of the
concurrent hitting time and the bound on it derived in
\cite{kempe_comp}. In the next section, this is redefined in terms
of the residual probability $1-p$ and plotted against the hitting
time defined in the previous section.

If we think of quantum walks as a possible route to new algorithms,
then the concurrent hitting time corresponds to the time needed to
find a solution with probability greater than p.  The definition of hitting
time used in this paper corresponds more to a typical running time
for the algorithm.  Both definitions could prove useful for particular purposes.

\section{Numerical results}

\subsection{Results for the Grover coin}

\begin{figure}[t]
\begin{center}
\includegraphics[scale=0.45]{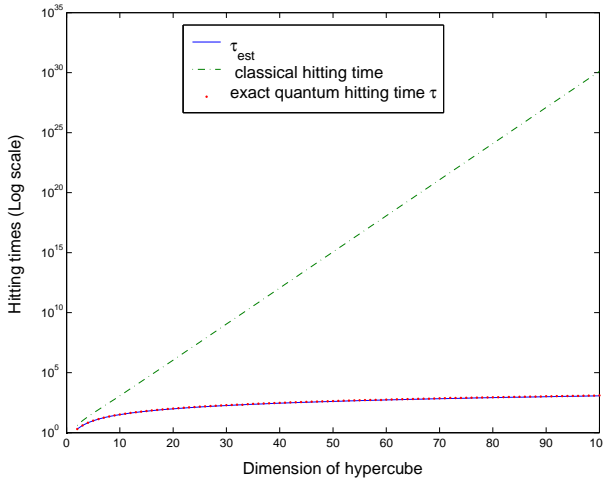}
\end{center}
\caption{\bf{Hitting times of Classical and Quantum walks
(semi-log scale)}} \label{htQandC}
\end{figure}

\begin{figure}[h]
\begin{center}
\includegraphics[scale=0.45]{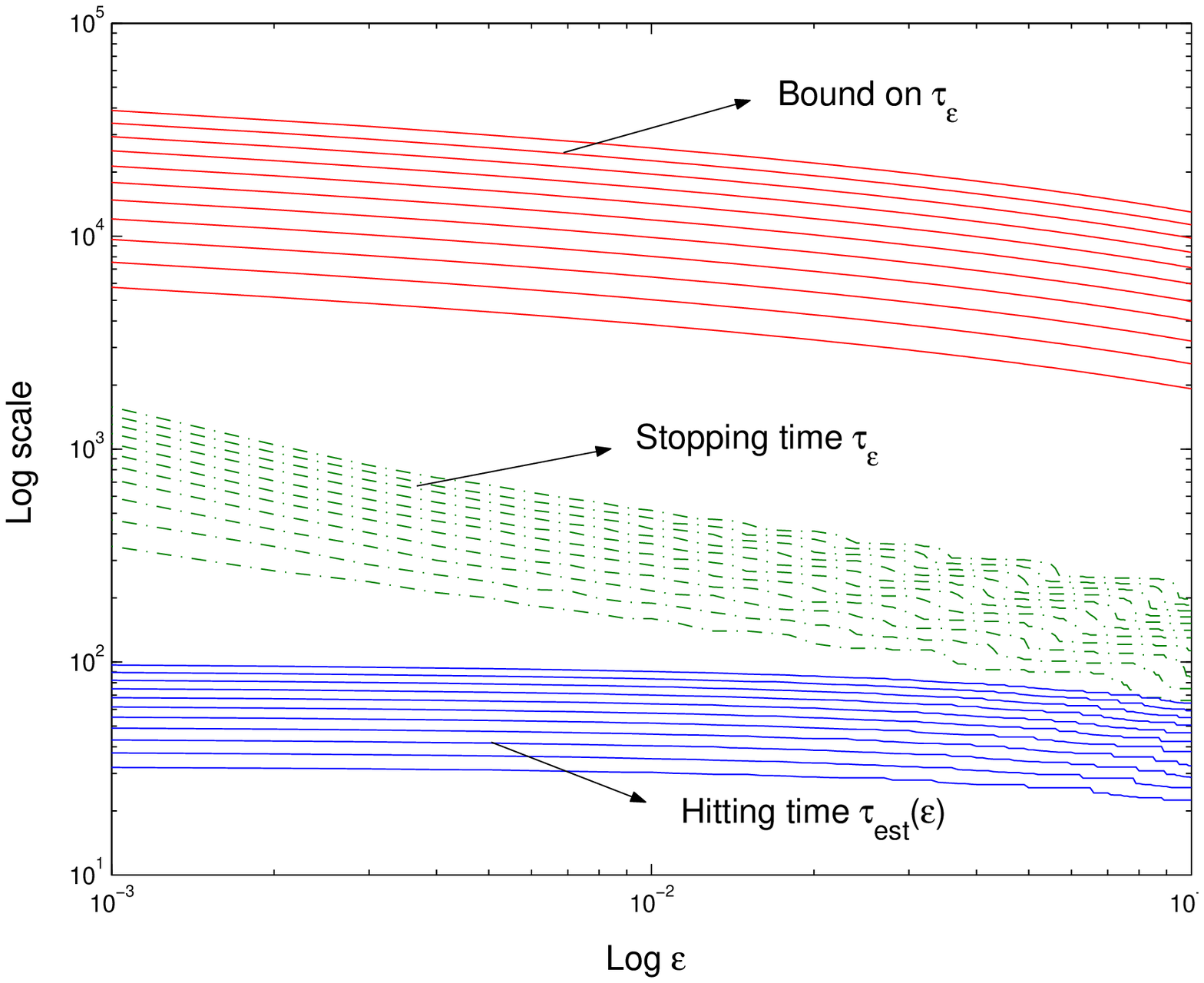}
\end{center}
\caption{\bf{Comparison of Quantum hitting time, stopping time and
the bound on the stopping time for dimensions 10 to 20(log scale)}}
\label{epsilon10_20}
\end{figure}

\begin{figure}[h]
\begin{center}
\includegraphics[scale=0.45]{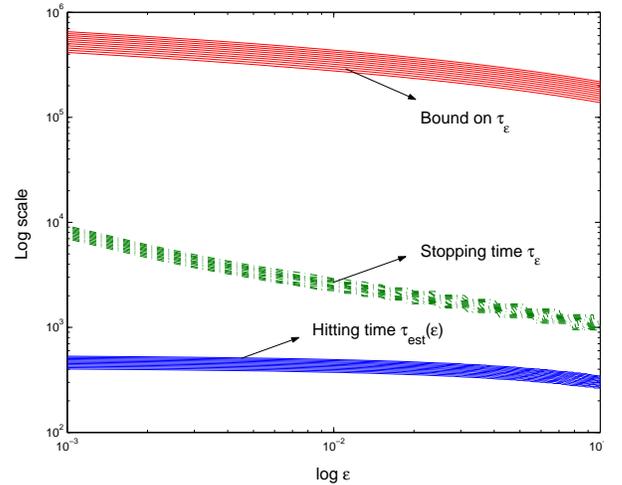}
\end{center}
\caption{\bf{Comparison of Quantum hitting time, stopping time and
the bound on the stopping time for dimensions 50 to 60(log scale)}}
\label{epsilon50_60}
\end{figure}

We calculated the hitting time by evaluating the above expression
(\ref{closed_form_tau}) in Matlab, and comparing the result to
a lower bound obtained by iterating the quantum walk for a large
number of steps.  Because of the multiple tensor products,
the size of the matrices $\mathcal{N}$ and $\mathcal{Y}$
is $(2^n n)^2\times(2^n n)^2$.

We can make this computation more tractable in the case where
the coin-flip unitary is the {\it Grover coin} $\hat{G}$
defined in (\ref{Grover_matrix}), and the specific starting
state is $\rho=|0\ra\la0| \otimes |\phi\ra\la\phi|$,
where $|\phi\ra=\frac{1}{\surd{n}}\sum_{i=1}^n|i\ra$.  It was shown by
Shenvi et al. \cite{shenvi} that for the walk on the hypercube
with this initial state and the Grover coin, the state remains always
in a $2n$-dimensional subspace, where the
walk is on a line with $n+1$ points.  This is rather similar to the
simplification made in the classical case, when we kept track
only of the Hamming weight of the current vertex.
With this simplification, the operators in (\ref{closed_form_tau})
reduce to $(2n)^2\times(2n)^2$
matrices, which makes explicit calculations possible even for
high dimensional hypercubes.

We write a set of basis states for this subspace as
$|R,0\ra,|L,1\ra,|R,1\ra,\dots,|R,n-1\ra,|L,n\ra$, where the first label
says whether the state is ``right-going'' or ``left-going,'' and the
second label gives the Hamming weight of the state.
The initial state is $|R,0\ra$, and the final state is $|L,n\ra$.
(Note that there are no states $|L,0\ra$ or $|R,n\ra$.)  Restricted
to this subspace, the $\hat{S}$ and $\hat{C}$ matrices become
\begin{equation}\label{S}
S=\sum_{x=0}^n|R,x\ra\la L,x+1|+|L,x+1\ra\la R,x|
\end{equation}
\begin{eqnarray}\label{C}
C|L,x\ra=-cos\omega_x|L,x\ra+sin\omega_x|R,x\ra \nonumber,
\\
C|R,x\ra=sin\omega_x|L,x\ra+cos\omega_x|R,x\ra,
\end{eqnarray}
where $\cos\omega_x = 1-2x/n$.  We see that for the walk in this
subspace, the coin flip is no longer independent of the position; this
is quite analogous to the reduction of the classical walk to the Hamming weight,
in which the probabilities favor walking towards $x=n/2$.

Now, going back to the definition of the hitting time as a sum of
series in Eq.~(\ref{ht}) where the probability is given by
Eq.~(\ref{probEqn}), we can define the $\epsilon$-stopping time
($\tau_{\epsilon}$) as the time step $T$ at which the total
probability $\sum_{i=1}^Tp(i)>1-\epsilon$, where $p(i)$ is as in
Eq.~(\ref{probEqn}).  We calculate an estimate of the hitting time
($\tau_{est}(\epsilon)$ as a function of $\epsilon$ by summing the
series in Eq.~(\ref{ht}) up to $\tau_{\epsilon}$. Figure
\ref{htQandC} shows the classical and quantum walks on the
hypercube for 100 dimensions. The exact hitting time $\tau$
calculated by computing the expression in equation (\ref{hittime})
is plotted as the dotted line and $\tau_{est}(\epsilon)$ for $\epsilon=0.001$ is plotted as the solid line. These two lines almost coincide in the graph. The hitting time on the Y-axis is plotted in log
scale. We can see that hitting time for the quantum walk is a low
order polynomial whereas the classical walk is exponential.
There is a very dramatic speed-up in the quantum case.

Figure \ref{epsilon10_20} plots the quantum hitting time, the
$\epsilon$-stopping time and the bound on the $\epsilon$-stopping
time (obtained in \cite{kempe_comp}) against $\epsilon$ for
dimensions from 10 to 20. Both the axes are in log scale. Figure
\ref{epsilon50_60} plots the same for dimensions from 50 to 60. It
can be seen from these two figures that the $\tau_{\epsilon}$ and
the bounds on it become less tight for higher dimensions.  Clearly,
the average walk ends much faster than these bounds might suggest.

\subsection{Results for the DFT coin}

The hitting time for the hypercube using the DFT coin, by contrast
to the Grover coin case considered above, can actually be infinite. For n=4 we
will demonstrate that for the same initial condition as with the Grover coin, the
hitting time for a quantum walk using the DFT coin is infinity.
This is because there exist eigenvalues of the evolution
operator $\mathcal{\hat{U}}$ whose eigenvectors have an overlap
with the initial state, but have no overlap with the final vertex
for any state of the coin.

Suppose there are states $|\phi\ra$ which have no overlap with the
final vertex,  $\la\phi|\hat{P}_f|\phi\ra = 0$, and which are eigenstates
of the evolution operator $\hat{U}$:   $\hat{U}|\phi\ra = \exp(i\theta)|\phi\ra$.
If the system is in the state $|\phi\ra$, clearly there is no probability to
ever detect the particle in the final vertex.  
Let $\mathcal{\hat{P}}$ be a projector onto all such states $|\phi\ra$.
Then $\mathcal{\hat{P}}\hat{P}_f = \hat{P}_f\mathcal{\hat{P}} = 0$, and
$[\mathcal{\hat{P}}, \hat{U}] = 0$.  One can write the initial state as a
superposition of vectors in this subspace and its orthogonal complement:
\begin{equation}
|\Psi\ra=\mathcal{\hat{P}}|\Psi\ra+(\mathcal{\hat{I}}-\mathcal{\hat{P}})|\Psi\ra .
\end{equation}
Any state that begins in the subspace selected by $\mathcal{\hat{P}}$
will remain there for all time, and any
state in the orthogonal complement will stay there; this follows from the fact
that the projectors commute with both the unitary transformation $\hat{U}$
and the measurement operator $\hat{P}_f$.  As one starts the walk, the
probability that the particle never reaches the final state is
$\la\Psi|\hat{U^{\dag}}^t\mathcal{\hat{P}}\hat{U}^t|\Psi\ra$,
which is $\la\Psi|\mathcal{\hat{P}}|\Psi\ra$.

In order for this probability to be nonzero, there must be eigenstates of
the unitary evolution operator $\hat{U}$ which have no amplitude for
the final vertex.  We can readily demonstrate this for the hypercube with
the DFT coin.  Consider the 4-dimensional hypercube.  Numerically diagonalizing
the evolution operator $\hat{U}$ given by (\ref{evolution_op}), we find it has $i,-i,1$ and
$-1$ among its eigenvalues, each with a degeneracy of 8.
Since the subspace corresponding to the final vertex is 4-dimensional,
it is clearly possible to construct a superposition of eigenvectors of any of these
eigenvalues so that it has no overlap with the final vertex in any
coin state.  For each of the four degenerate eigenvalues we can construct
a 4-dimensional subspace of eigenvectors with no overlap with the final
vertex, giving a 16-dimensional space for all such eigenvectors.
By numerically constructing an orthonormal basis for this space, we can
find an expression for the projector $\mathcal{\hat{P}}$ and measure
its overlap with the initial state.

We considered in particular the
initial state where the particle was located at the $|00\dots0\ra$ vertex
and the coin is in an equal superposition of basis states $|1\ra,\ldots,|n\ra$.
For the hypercube with $n=4$ and the given initial state,
the probability $\la\Psi|\mathcal{\hat{P}}|\Psi\ra$ is
$0.4286$, which exactly matches the total probability to never
hit the final node after a large number of iterations in our
numerical simulations.  Thus, the probability is close to half that the
particle never reaches the final state and the hitting time
becomes infinity.

This demonstrates a property of quantum walks
not seen in their classical counterparts:  for certain initial
conditions, there is a nonzero probability that the particle never
reaches the final state, even though the initial and final states
of the graph are connected.  For a quantum walk with substantial
degeneracy, this phenomenon is likely to be generic.  It might
be possible to make the hitting time finite by choosing an appropriate
initial condition---clearly this happens for the Grover coin---but for
some coins this may require an initial condition which is not
localized on one vertex.  From our simulations, it it seems that for
higher dimensions the DFT coin behaves similarly to $n=4$.
For example, for $n=5$, our simulations show that the
probability to hit the final node increases slowly but does not
reach 1 even after many time steps. This could be due to the
fact that the final vertex has no overlap with some eigenvectors
of the evolution operator (as for $n=4$), and additionally that
it overlaps very little for some other eigenvector.  This would make
the probability increase slowly but never reach 1.

\subsection{Results for a quantum walk on a distorted hypercube}

If, as seems likely, the dramatic speed-ups (and slow downs) of
quantum walks over their classical counterparts depend on the
symmetry of the graph, it should be instructive to see the effect
of deviations from that symmetry.  In this section, we look at
results for the measured walk using the Grover coin on a
distorted hypercube. The distorted hypercube is defined by
constructing the usual hypercube, and then switching two
of the connections.  Pick 4 vertices which form a face--for example,
$(0\dots00),(0\dots01),(0\dots10),(0\dots11)$. Calling these vertices
$A,B,C,D$ for short, we distort the hypercube by connecting $A$ to $D$
and $B$ to $C$, and removing the edges between $A$ and $B$ and
between $C$ and $D$. This is still a regular graph, and the same
quantum walk can be used without having to redefine the evolution
operator.  Unlike the usual hypercube, it is no longer a bipartite graph,
and the walk can no longer be reduced to a walk in Hamming weight.

\begin{figure}[t]
\begin{center}
\includegraphics[scale=0.45]{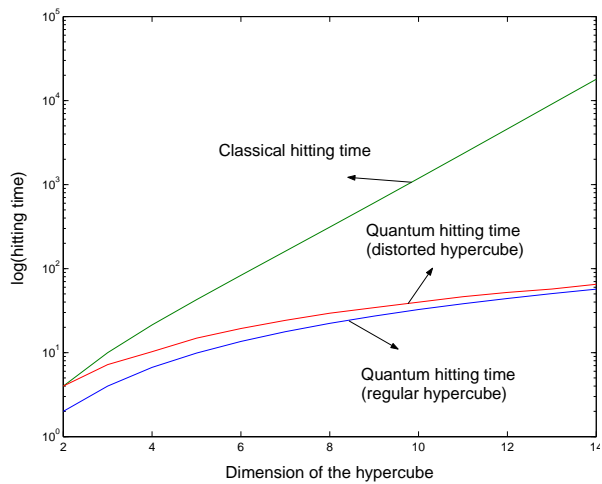}
\end{center}
\caption{\bf{Comparison of hitting times on the regular and
distorted hypercubes}} \label{skewed}
\end{figure}

Figure \ref{skewed} plots the hitting time of a quantum walk on a
distorted hypercube together with that of a classical walk and a
quantum walk on regular hypercubes for comparison. The hitting
time for a quantum walk on a distorted hypercube is more than that
of a quantum walk on a regular hypercube, but still is much
smaller than the hitting time of a classical walk. In fact, as the
dimension increases one can see that the hitting times of the
quantum walk on the distorted and regular hypercubes converge
towards each other.  This presumably reflects the fact that for
higher dimensions the symmetry is mostly unchanged.

\section{Conclusions}

We have examined the definition of hitting time for measured quantum
walks and analyzed some of its properties. We used this definition to
obtain an expression for hitting time which is valid on any general
graph as long as the unitary evolution operator $\hat{U}$ of the walk
is defined. We simulated this hitting time for
a measured quantum walk using the Grover coin and compared it to
the classical hitting time and to the bounds obtained on it; the quantum
hitting time is exponentially smaller than the classical
hitting time. We also showed that the bounds on the hitting time
obtained in \cite{kempe_comp} become less tight as the
dimension increases.

However, simply making a walk quantum does not guarantee a speed-up
over the classical case.  We demonstrated that the hitting time for quantum
walks can depend sensitively on the initial condition, unlike classical walks.
For certain initial states, the DFT walk can have {\it infinite} hitting
time, a phenomenon not possible in classical random walks. This
dependence on the initial state varies with the coin used, since
for the same initial state the Grover walk has a polynomial
hitting time. This infinite hitting time is directly related to the
degeneracy of the eigenvalues of the evolution operator.
If the evolution operator is highly degenerate, then it is very likely
that there exist initial states which give infinite hitting times.

While the exact cause of the speed-up in quantum hitting time is not
completely clear, it seems very likely that the symmetry of the graph
plays a major role in both the speed-up and slow down of the quantum
walk. In the faster quantum walk, the different paths
leading to the final vertex interfere constructively, enhancing the probability
of arrival; paths which lead to ``wrong'' vertices interfere destructively,
reducing the probability of meandering around in the graph for long times.
Unlike a classical random walk, the quantum walk is sensitive to the presence
of a global symmetry which is not apparent at a purely local level.  This
phenomenon probably leads to the speed-up of the continuous-time
quantum walk on the glued-trees graph as well \cite{childs2}.

However, this same symmetry is undoubtedly the culprit in the {\it slow-down}
observed for the DFT walk.  The existence of states which never arrive at
the final vertex is made possible by the degeneracy of the evolution operator
$\hat{U}$---a degeneracy which arises due to the symmetry of the graph.
The existence of states which never arrive at the final vertex can also be seen
as an interference effect, only in this case the interference of paths which lead
to the final vertex is {\it destructive}:  all amplitude to make a transition to
the final vertex cancels out.

This interpretation is supported by the quantum walk on the distorted
hypercube. We observe that the hitting time is worse than that of
the usual hypercube, but still much smaller than that of a
classical walk. The curve of the hitting time on the distorted
hypercube seems to converge slowly to that of a quantum walk on
the regular hypercube. This is probably because the distortion
(described in section 4) is very mild. As the dimension grows, and
with it the number of edges and vertices, this distortion has less
effect on the overall symmetry.

We hasten to add that symmetry of the graph is not the sole reason
for speed-ups in quantum walks.  A polynomial speed-up has been
demonstrated in the quantum walk versions of the search of an 
unstructured database \cite{shenvi} and the element distinctness
problem \cite{ambainisApp}.  However, the dramatic exponential speed-ups
have all been demonstrated in highly symmetric graphs.

This suggests that the most promising direction to look for new algorithms
based on quantum walks is for problems which possess a global symmetry,
but for which this symmetry is not apparent at the level of
individual candidate solutions.  Yet even for such a problem, care will
have to be taken if quantum mechanics is to serve as a blessing and not
as a curse.

\end{document}